# UNDERSTANDING ORAL PRESENTATIONS THROUGH AUDIENCE GAZE: AN EXPLAINABLE LEARNING ANALYTICS APPROACH

Pablo Villegas[1], Alvaro Becerra[1], Ruth Cobos[1] and Xavier Ochoa[2]
[1]*Universidad Autónoma de Madrid*
*Ciudad Universitaria de Cantoblanco, 28049 Madrid, Spain*
[2]*New York University*
*New York, NY, USA*

**ABSTRACT**

Oral presentation assessment is typically based on human rubrics, which provide structured evaluation but limited evidence about how a presentation is received by the audience. This study investigates whether eye-tracking data from an observer can provide interpretable information related to teacher-based assessment of face-to-face oral presentations. The dataset comprises 71 presentations from two undergraduate engineering cohorts. Observer gaze was mapped to Areas of Interest, and visual-attention descriptors were transformed into stable and temporal session-level representations. Regression and high-vs-low classification models were evaluated for an attention-related composite score (Attention Core) and the overall Teacher Rubric Mean. Under leave-one-out validation, regression achieved $R^2$=0.330 for Attention Core and $R^2$=0.383 for Teacher Rubric Mean, while high-vs-low Attention Core classification reached a balanced accuracy of 0.802 and ROC AUC of 0.909. Explainability analysis highlighted presenter-directed attention, speaker–screen balance, gaze revisits, transition dynamics, and presentation-specific temporal segments. These findings suggest that observer eye tracking can complement human assessment with interpretable evidence about how oral presentations are visually received, supporting future feedback-oriented applications.

## 1. INTRODUCTION

Oral presentation skills are essential in higher education, but their assessment often relies on human rubrics that, while structured, remain subject to evaluator variability and provide limited objective evidence about how presentations are received (Morreale & Pearson, 2008).

Multimodal Learning Analytics (MMLA) addresses some of these limitations by combining heterogeneous sources, such as biometric signals and behavioral and contextual data, to provide a more holistic understanding of the learning process (Becerra, Cobos, & Lang, 2026; Giannakos et al., 2022). Among these sources, eye tracking has been widely used in educational research to study visual attention and how learners allocate attention across elements of a learning environment to support personalized learning and feedback (Golrang & Sharma, 2026; Soares et al., 2023)

In oral presentation contexts, MMLA has been applied mainly through Automated Oral Presentation Feedback (OPAF) systems, which use multimodal evidence to support the development of presentation skills (Ochoa, 2022). Systems such as Presentation Trainer (Schneider et al., 2015), RAP (Ochoa et al., 2018), and OpenOPAF (Ochoa & Zhao, 2024) provide automated feedback based on features related to speech, posture, gestures, facial behavior, and visual support. However, despite their different feedback strategies, these approaches have predominantly adopted a presenter-centred perspective, analyzing how the presentation is delivered rather than how it is received by the audience.

Against this background, this study adopts an observer-centred perspective by analysing eye-tracking data from an external audience member during face-to-face oral presentations. The contribution is twofold: first, to examine whether interpretable visual-attention descriptors are associated with attention-related and broader teacher evaluations; and second, to analyse how stable and temporal gaze representations, together with model explainability, can provide evidence about both what visual-attention patterns are relevant and when they occur during the presentation. Rather than replacing human assessment, the aim is to complement rubric-based evaluation with interpretable evidence that can support more specific and temporally grounded feedback.

## 2. STUDY CONTEXT

The study comprises 71 individual face-to-face oral presentations delivered by undergraduate engineering students enrolled in the Engineering and Society (Ingeniería y Sociedad) course at Universidad Autónoma de Madrid (UAM). The dataset combines two academic cohorts collected in the same educational setting: 39 presentations from the 2024/2025 cohort, which are part of the SOPHIAS multimodal oral-presentation dataset (Becerra, Cobos, & Daza, 2026), and 32 additional presentations from the 2025/2026 cohort. In both cohorts, data collection followed the same acquisition procedure described for SOPHIAS. Each presentation includes synchronized external-observer eye-tracking data, the scene video recorded by the eye-tracking glasses, and rubric-based human evaluations.

The presentation scenario involved three main roles. First, the presenter was the student delivering an individual oral presentation supported by projected slides. Second, the teacher assessed the presentation using a predefined rubric. Third, an external observer attended the presentation as an audience member while wearing eye-tracking glasses. The observer did not provide the evaluation target. A different external observer participated in each cohort. All sessions were collected under informed-consent procedures and participation was voluntary.

The teacher assessed each oral presentation through the AICoFe system (Becerra et al., 2027) using a 15-item rubric covering three main dimensions of oral presentation performance: content, slide design, and presenter performance. The rubric included items related to presentation opening, content development, presentation closing, bibliography and references, text readability, relevance of visual elements, visual balance, voice and tone, body expression, visual contact with the audience, use of complementary resources, ability to capture attention, autonomy during the presentation, handling of questions, and topic mastery. Each item was rated on a five-point Likert scale ranging from 1 (very poor) to 5 (excellent).

In this study, two target variables were derived from these teacher evaluations. Attention Core was computed as the mean of the rubric items Ability to Capture Attention, Visual Contact with the Audience, and Body Expression, representing the dimensions most closely related to attention and non-verbal communication. Teacher Rubric Mean was calculated as the mean of the 15 rubric-item scores and represents a broader assessment of overall presentation performance.

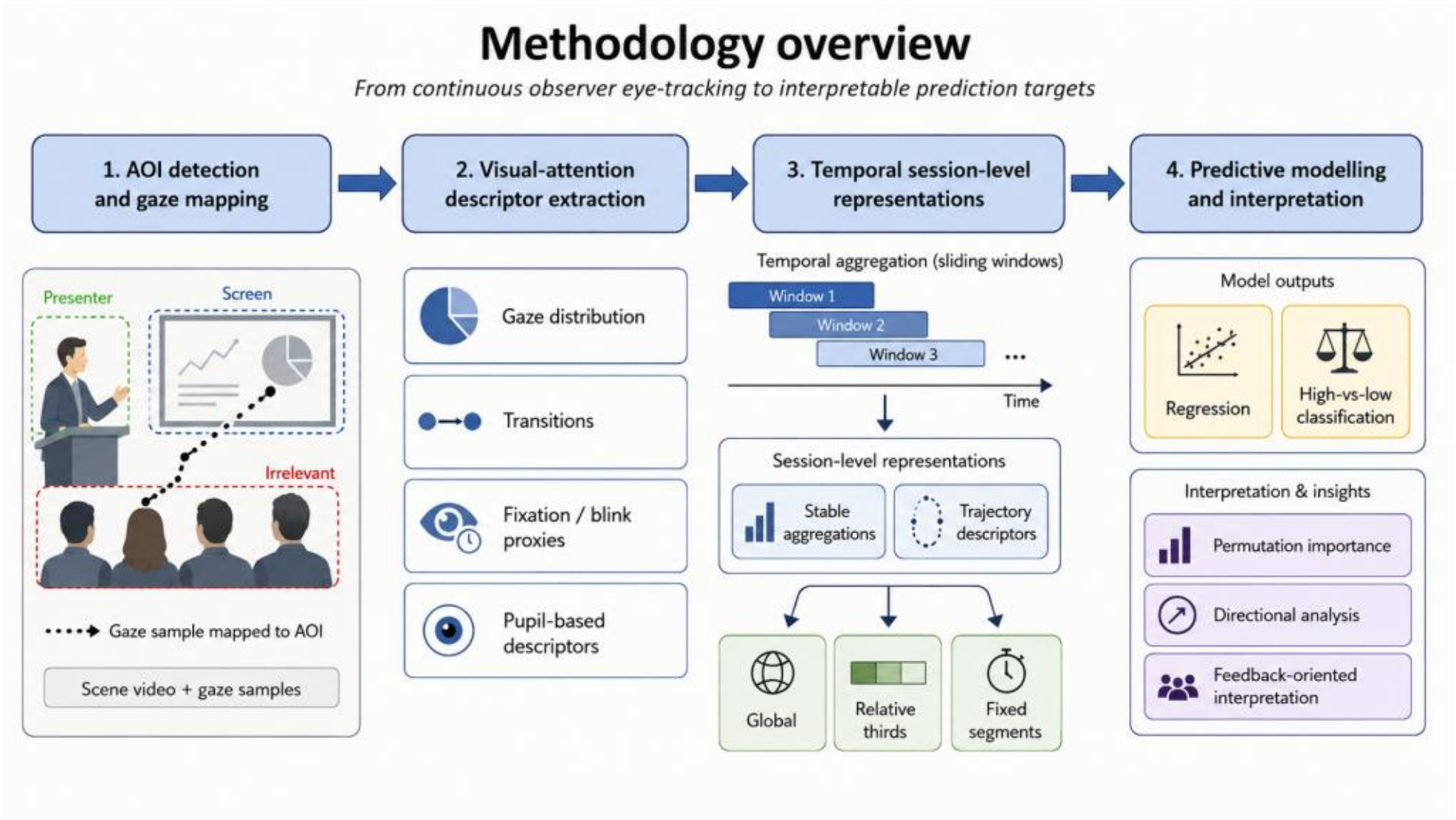

Figure 1. Overview of the proposed observer-centred eye-tracking methodology

# 3. METHODS

The proposed methodology transforms the external observer's continuous eye-tracking recordings into interpretable session-level representations that can be related to the human-evaluation targets. The processing pipeline comprises four stages: semantic Area-of-Interest (AOI) detection and gaze mapping, extraction of visual-attention descriptors, construction of temporal session-level representations, and predictive modelling and interpretation, as summarized in Figure 1.

## 3.1 AOI Detection and Gaze Mapping

To provide a semantic interpretation of the observer's gaze, the scene video recorded by the eye-tracking glasses was processed offline to identify the main visual regions of the presentation environment. A custom detector based on YOLOv8 was trained on manually annotated scene frames to detect the presenter and the projected screen. YOLOv8n, YOLOv8s, and YOLOv8m were compared, and YOLOv8n was selected as the final model because it provided an appropriate trade-off between detection performance and computational cost, achieving a precision of 0.964, recall of 0.970, mAP50 of 0.980, and mAP50–95 of 0.889 (Redmon et al., 2016; Swathi & Challa, 2024).

The resulting bounding boxes were used to map each valid gaze sample to a semantic AOI as shown in Figure 2. Gaze coordinates falling within the presenter or screen bounding boxes were assigned to the corresponding AOI, while valid samples outside both regions were labelled as irrelevant.

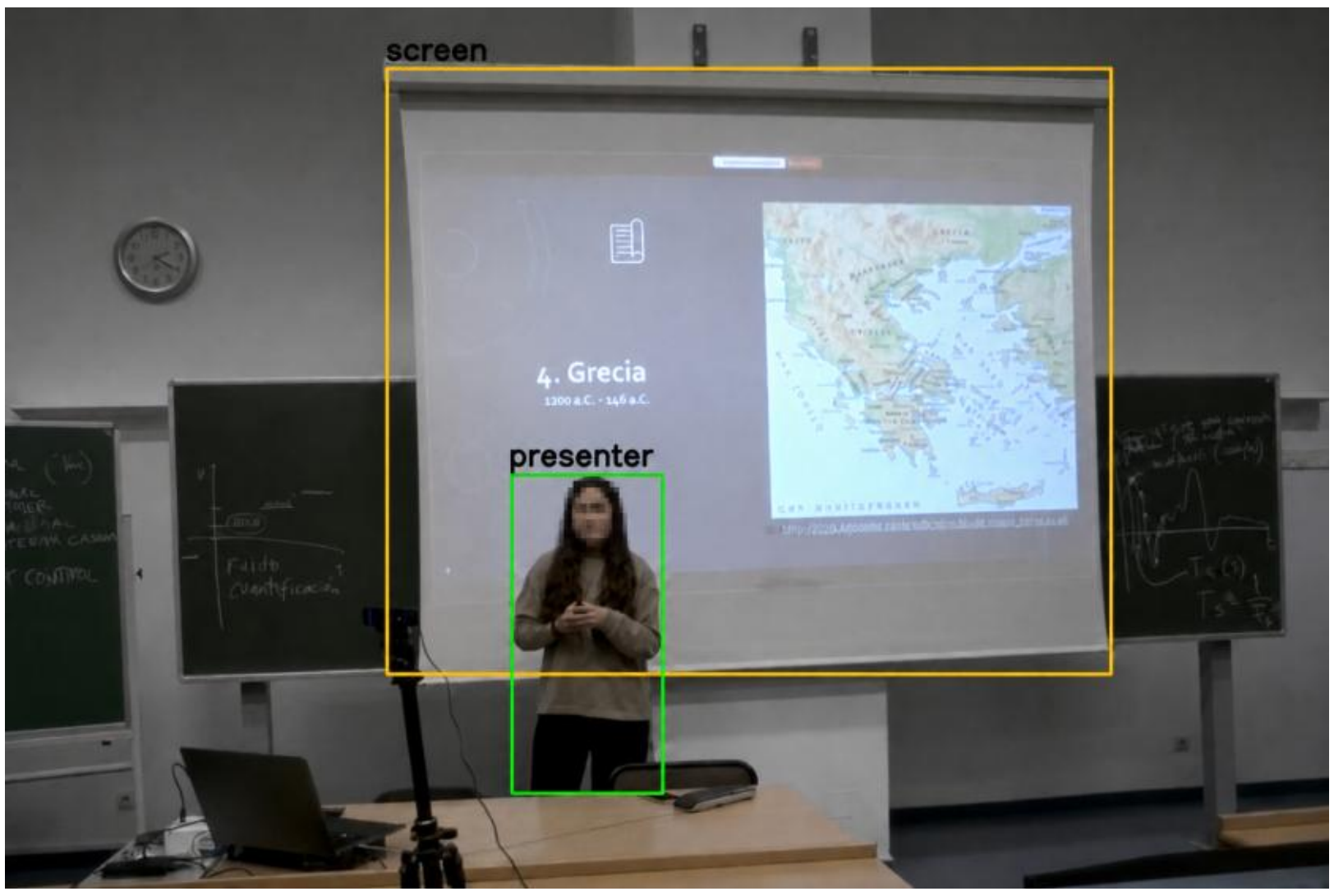

Figure 2. Example of semantic AOI detection in the eye-tracker scene video

## 3.2 Visual-Attention Descriptors

Visual-attention descriptors were computed over multiple temporal window-overlap configurations to reduce sample-level variability while retaining local changes in attention throughout the presentation. The evaluated configurations included different window lengths and overlap settings (5, 10, 15, 20, and 30-second windows, combined with no overlap, 25% overlap, 50% overlap, 1-second overlap, and 2-second overlap), allowing the models to compare more local and more smoothed representations of observer visual behaviour. After preprocessing, AOI mapping, and feature cleaning, 27 eye-tracking descriptor trajectories were retained for session-level representation.

The window-level eye-tracking descriptors were organized into four main families. Gaze-distribution descriptors characterized where the observer allocated visual attention and included dwell time on the presenter, screen, and irrelevant regions, total dwell time on relevant AOIs, and speaker–screen balance and ratio. Gaze-transition descriptors characterized how attention moved between these regions and included AOI switching rates, relevant and irrelevant switching rates, revisit rate, transition entropy and its variants, and gaze-step amplitude. Fixation and blink descriptors included proxy measures of mean fixation duration on the presenter and screen, blink rate, and mean blink duration. Finally, pupil-based descriptors included the mean, standard deviation, and temporal variation of normalized pupil size, the proportion of high pupil-activity samples, and IPA- and LHIPA-based measures. These descriptors are grounded in established eye-tracking measures of AOI-based attention, fixation behaviour, gaze transitions, and pupil activity (Duchowski, 2017; Duchowski et al., 2018; Holmqvist et al., 2011; Krejtz et al., 2015).

## 3.3 Temporal and Session-Level Representation

Since eye-tracking provides a continuous temporal signal whereas teacher evaluation is defined once per presentation, the window-level descriptors were transformed into session-level representations. Each of the 27 selected visual-attention descriptors was considered as a temporal trajectory describing its evolution throughout a presentation. Two complementary representations were extracted from each trajectory. First, stable aggregation features summarized its overall behaviour using eight statistics: mean, standard deviation, median, minimum, maximum, and the 25th, 75th, and 90th percentiles. This resulted in 216 features for the global session representation. Second, each trajectory was characterized using a set of 32 temporal descriptors capturing its dynamic behaviour, including accumulated positive and negative variation, velocity-, acceleration-, and jerk-related measures, temporal locations of extrema, sign changes, local maxima, range, skewness, and kurtosis. This family of temporal descriptors has previously been used in biometric and attention-analysis tasks to transform dynamic signals into fixed-length representations (Becerra, Daza, et al., 2025; Daza et al., 2024; Fierrez-Aguilar et al., 2005) (Fierrez-Aguilar et al., 2005; Daza et al., 2024; Becerra et al., 2026). Applying the 32 descriptors to the 27 eye-tracking trajectories produced 864 temporal features per presentation. A third representation combined stable and temporal features, resulting in 1,080 global features.

To analyze when relevant visual-attention patterns occurred, the same representations were also extracted from different temporal regions of the presentation. In addition to the global session, the presentation was divided into beginning, development, and final segments. Relative thirds and fixed-length beginning/end configurations of 30, 45, 60, 90, and 120 seconds were evaluated.

## 3.4 Predictive Modelling and Model Interpretation

Two prediction tasks were considered. For regression, the continuous Attention Core and Teacher Rubric Mean scores were predicted. A mean dummy regressor, Ridge Regression, Elastic Net, and Random Forest were evaluated. Missing values were median-imputed and univariate feature selection was performed using SelectKBest with the $F$-regression score, retaining up to 100 features. Linear models additionally included feature standardization.

For a high-vs-low classification, the continuous targets were converted into two clearly separated classes using their lower and upper tertiles, while presentations in the intermediate range were excluded. For Attention Core, presentations with scores $\leq 6.67$ were labelled low and those with scores $>8.00$ were labelled high, resulting in 50 presentations (28 low, 22 high). For Teacher Rubric Mean, scores $\leq 6.67$ were labelled low and scores $>8.17$ high, resulting in 48 presentations (24 low, 24 high). The classification experiments considered linear and non-linear models, including Logistic Regression, linear SVM, Random Forest, and Histogram-based Gradient Boosting (HGB).

Model selection was performed using shuffled 5-fold cross-validation, comparing combinations of temporal configuration, feature representation, and model. Regression configurations were selected according to cross-validated $R^2$, with MAE as a secondary criterion, and the selected configuration was subsequently evaluated using Leave-One-Out Cross-Validation (LOOCV). Hyperparameters were fixed beforehand rather than optimized through an additional search. The best-performing models were further analysed from an explainability perspective. Permutation importance with 30 repetitions was used to quantify the contribution of each feature, using $R^2$ as the scoring function for regression and balanced accuracy for classification. To facilitate interpretation of the high-dimensional representations, importance values were aggregated by base eye-tracking descriptor and by temporal segment (global, beginning, development, and final).

In addition, a weighted Spearman correlation analysis was performed to determine the direction of the association between the most relevant descriptors and the evaluation targets, distinguishing whether higher feature values were associated with higher or lower evaluation scores. These analyses were used to identify the visual-attention patterns and presentation segments most influential in the selected models.

# 4. RESULTS

## 4.1 Model Results

Table 1 summarizes the best regression configuration obtained for each target. Attention Core was better modelled using stable aggregation features, whereas Teacher Rubric Mean benefited from temporal trajectory features.

Table 1. Best regression configurations and performance metrics for each prediction target

| Target | Window | Overlap | Temporal configuration | Features | Model | 5-fold $R^2$ | LOOCV $R^2$ | MAE | Pearson | Spearman |
|---|---|---|---|---|---|---|---|---|---|---|
| Attention Core | 10s | 25% | 45 s | Stable aggregation | Ridge | 0.373 | 0.330 | 1.351 | 0.600 | 0.615 |
| Teacher Rubric Mean | 10s | 2s/20% | 120 s | Temporal trajectory | Ridge | 0.293 | 0.383 | 0.890 | 0.629 | 0.597 |

Table 2 compares the three session-level representations using their best 5-fold cross-validated $R^2$, showing that the combined representation did not improve performance for either target.

Table 2. Comparison of feature representations using 5-fold cross-validated $R^2$

| Target | Stable aggregation | Temporal trajectory | Combined |
|---|---|---|---|
| Attention Core | 0.373 | 0.235 | 0.299 |
| Teacher Rubric Mean | 0.165 | 0.293 | 0.243 |

Table 3 reports the final high-vs-low classification results under LOOCV, showing that the strongest discrimination was obtained for Attention Core.

Table 3. High-vs-low classification results under LOOCV

| Target | N (Low/High) | Window | Overlap | Temporal configuration | Features | Model | Accuracy | Bal. Acc. | F1 | ROC AUC |
|---|---|---|---|---|---|---|---|---|---|---|
| Attention Core | 50 (28/22) | 10s | 2s/20% | Thirds | Stable | HGB | 0.800 | 0.802 | 0.783 | 0.909 |
| Teacher Rubric Mean | 48 (24/24) | 5s | 50% | 90 s | Stable | Linear SVM | 0.667 | 0.667 | 0.667 | 0.701 |

## 4.2 Explainability Results

Permutation importance revealed different explanatory patterns across the selected models. Table 4 reports the five most relevant base eye-tracking descriptors after aggregating the importance of all derived features belonging to each descriptor.

Attention Core relied mainly on presenter-directed attention, gaze recurrence, presenter–screen balance, fixation stability, and pupil-related descriptors. In contrast, Teacher Rubric Mean was more strongly characterized by gaze-transition dynamics and changes in the organization of visual attention.

Temporal aggregation of permutation importance showed that feature relevance was also distributed differently across presentation phases (Table 5).

Table 4. Most relevant base eye-tracking descriptors by target and task

| Target | Task | Most relevant base descriptors |
|---|---|---|
| Attention Core | Regression | revisit rate (0.177); dwell speaker (0.171); speaker–screen balance (0.130); pupil LHIPA mean (0.097); pupil z-score delta (0.078) |
| Attention Core | Classification | pupil LHIPA mean (0.058); fixation duration speaker proxy (0.045); pupil z-score std (0.032); revisit rate (0.029); pupil z-score mean (0.026) |
| Teacher Rubric Mean | Regression | revisit rate (0.194); collapsed transition entropy (0.085); transition entropy (0.052); irrelevant switching rate (0.051); speaker–screen balance (0.033) |

Table 5. Aggregated permutation importance by target, task, and presentation segment

| Target | Task | Global | Beginning | Development | Final |
|---|---|---|---|---|---|
| Attention Core | Regression | 0.341 | 0.116 | 0.331 | 0.084 |
| Attention Core | Classification | 0.010 | 0.011 | 0.056 | 0.110 |
| Teacher Rubric Mean | Regression | 0.090 | 0.256 | −0.012 | 0.042 |

The Attention Core regression model concentrated its relevance in the global and development representations, whereas its high-vs-low classifier relied more strongly on the final part of the presentation. For Teacher Rubric Mean, the beginning contained the largest aggregated importance.

Finally, the directional analysis showed a consistent positive pattern for Attention Core regression: higher presenter dwell time, speaker–screen balance, revisit rate, pupil LHIPA mean, and pupil z-score variation were associated with higher Attention Core scores. The Attention Core classifier similarly associated stable presenter-directed gaze and recurrent visual engagement with the high-scoring class, although pupil z-score mean showed the opposite direction. For Teacher Rubric Mean, the relevant descriptors showed mixed directions: collapsed transition entropy was positively associated with the target, whereas transition entropy, revisit rate, pupil LHIPA mean, and irrelevant switching rate showed negative associations.

# 5. EDUCATIONAL INTERPRETATION

The predictive results indicate that observer eye-tracking is most informative when the evaluation target is closely related to visual attention and non-verbal communication. Although the regression performance was moderate, this should be considered in light of the complexity of predicting rubric-derived human evaluations from a single behavioural modality. Attention Core was better represented by stable summaries of gaze behaviour, whereas Teacher Rubric Mean benefited more from descriptors capturing how visual attention evolved throughout the presentation. This difference is consistent with the nature of the targets: Attention Core directly reflects attention capture, visual contact, and body expression, while Teacher Rubric Mean integrates additional dimensions that are not directly observable from gaze. Observer eye-tracking should therefore be interpreted primarily as evidence of how a presentation is visually received, rather than as a direct measurement of overall presentation quality.

The high-vs-low results reinforce this interpretation. Observer eye-tracking was particularly effective when distinguishing clearly different Attention Core profiles. This formulation is relevant for formative applications because clearly weak and strong attention-related profiles may provide more meaningful intervention points than small differences between neighbouring rubric scores.

The explainability analysis provides a more concrete interpretation of the visual behaviour captured by the models. For Attention Core, one of the most relevant descriptors was dwell speaker, defined as the proportion of valid gaze samples directed towards the presenter. Higher presenter-directed dwell was associated with higher Attention Core scores, suggesting that presentations receiving stronger attention-related evaluations tended to maintain the observer's gaze on the presenter. Speaker–screen balance measures whether attention is concentrated on one of the two main visual sources or distributed between presenter and slides. Its positive association suggests that stronger Attention Core profiles were not characterized simply by maximizing gaze towards the presenter, but by maintaining a balanced visual relationship between the presenter and the supporting material. Revisit rate, which measures how frequently the observer returns to previously visited AOIs, was also relevant. This may reflect recurrent re-engagement with the main presentation elements rather than prolonged disengagement from the scene. Finally, the speaker fixation-duration proxy captures approximately how long stable gaze periods towards the presenter are maintained. Together, these descriptors point to a pattern combining sustained presenter-directed attention, recurrent visual re-engagement, and coordinated attention between presenter and slides.

Teacher Rubric Mean showed a different and less direct pattern. Its most relevant descriptors were mainly related to gaze-transition dynamics, including revisit rate, transition entropy, and irrelevant switching rate. Transition entropy quantifies the variability and predictability of gaze transitions between AOIs: higher values indicate a more variable or less predictable sequence of visual shifts. Irrelevant switching rate specifically captures transitions involving regions outside the presenter and screen. The directional associations were mixed rather than consistently positive or negative. This suggests that broader teacher evaluation is not captured by a simple rule such as “more gaze activity is better”, but rather by how visual attention is organized and evolves across the presentation. For example, structured transitions between relevant AOIs may reflect effective coordination between presenter and visual support, whereas frequent transitions involving irrelevant regions may indicate moments in which the presentation was less successful at maintaining visual engagement.

Pupil-based variables also appeared among the relevant Attention Core descriptors. Pupil z-score mean, pupil z-score standard deviation, and pupil z-score delta summarize normalized pupil level, variability, and change, while LHIPA captures frequency-related dynamics of pupil activity. Their relevance suggests that pupil dynamics added complementary information about the observer's response.

At the temporal level, the segment-level analysis suggested that relevant information was not uniformly distributed across the presentation. Attention Core was mainly associated with global and development-stage patterns, whereas Teacher Rubric Mean showed stronger relevance at the beginning of the presentation. This suggests that different evaluation targets may depend on different temporal moments, supporting future work on temporally localized feedback.

As future work, these observer-centred patterns could be translated into temporally localized feedback for students, linking rubric-based evaluations with observable evidence about how the presentation was visually received. This could complement human-based feedback with data-based insights within frameworks such as MOSAIC-F (Becerra, Andres, et al., 2025), although the usefulness and pedagogical value of such feedback should be evaluated with teachers and students.

## 6. CONCLUSION

This study shows that external-observer eye tracking can provide complementary and interpretable evidence for the assessment of face-to-face oral presentations. The strongest results were obtained for attention-related evaluation, supporting the use of observer-centred visual behavior as a source of information about how a presentation is received rather than as a direct measure of presentation quality.

These findings support the integration of observer-centred evidence into future feedback-oriented systems, where it could complement human assessment with temporally grounded information about audience visual response. A main limitation is that each presentation was represented by the gaze of a single external observer, whose visual behaviour may be affected by individual factors such as interest, fatigue, or attentional variability. Future work should therefore incorporate synchronized eye tracking from multiple audience members to obtain a more representative measure of audience reception and to assess whether aggregating across observers improves the explanatory power of the models, particularly for broader evaluation targets that include dimensions not directly observable from gaze.

## ACKNOWLEDGEMENT

Support by projects: Cátedra ENIA UAM-VERIDAS en IA Responsable (NextGenerationEU PRTR TSI-100927-2023-2), M2RAI (PID2024-160053OB-I00, MICIU/FEDER), Emotional (UAM AVANZA 2026) and SNOLA (RED2022-134284-T). Alvaro Becerra is funded by a predoctoral contract (FPI) from the Comunidad de Madrid (PIPF-2024/COM-34288).